\documentclass{llncs}
\usepackage{tikz,tkz-graph,amsmath,amsfonts,amssymb,graphics,color,comment,graphicx,fullpage}
\usepackage{subcaption}
\usetikzlibrary{arrows.meta}
\usetikzlibrary{arrows}
\usetikzlibrary{arrows.meta}
\usetikzlibrary{decorations.pathmorphing}
\usetikzlibrary{decorations.markings}
\usetikzlibrary{calc}

\tikzset{
  circ/.style = {circle,draw,fill,inner sep=1pt},
  lcirc/.style = {circle,draw,fill,inner sep=4pt},
  scirc/.style = {circle,draw,fill,inner sep=.5pt},
  mcirc/.style = {circle,draw,fill,inner sep=2pt},
  invisible/.style = {circle,draw=none,inner sep=0pt,font=\tiny}
}

\newtheorem{observation}{Observation}
\newtheorem{Claim}{Claim}

\begin{document}


\title{Planar CPG graphs}

\author{Nicolas Champseix\inst{1}
\and Esther Galby\inst{2}
\and Bernard Ries\inst{2}}

\institute{\'Ecole Normale Sup\'erieure de Lyon, France
\and Department of Computer Science, University of Fribourg, Switzerland}


\maketitle

\begin{abstract} We show that for any $k \geq 0$, there exists a planar graph which is $B_{k+1}$-CPG but not $B_k$-CPG. As a consequence, we obtain that $B_k$-CPG is a strict subclass of $B_{k+1}$-CPG.
\end{abstract}


\section{Introduction}

A graph $G=(V,E)$ is a \textit{contact graph of paths on a grid} (or \textit{CPG} for short) if there exists a collection $\mathcal{P}$ of interiorly disjoint paths on a grid $\mathcal{G}$ in one-to-one correspondence with $V$ such that two vertices are adjacent in $G$ if and only if the corresponding paths touch;  if furthermore every path has at most $k$ bends (i.e. 90-degree turn at a grid-point) for $k \geq 0$, then the graph is \textit{$B_k$-CPG}. The pair $\mathcal{R} = (\mathcal{G}, \mathcal{P})$ is a \textit{CPG representation of $G$}, and more specifically a \textit{$k$-bend CPG representation of $G$} if every path in $\mathcal{P}$ has at most $k$ bends. It was shown in \cite{cpg} that not all planar graphs are CPG and that there exists no value of $k\geq 0$ for which $B_k$-CPG is a subclass of the class of planar graphs. In this note, we show that there exists no value of $k$ such that $B_k$-CPG contains the class of planar CPG graphs. More specifically, we prove the following theorem.

\begin{theorem}
\label{thm:unbound}
For any $k \geq 0$, there exists a planar graph in $B_{k+1}$-CPG $\setminus B_k$-CPG.
\end{theorem}

It immediately follows from the definition that $B_k$-CPG $\subseteq$ $B_{k+1}$-CPG but it was not known whether this inclusion is strict; Theorem \ref{thm:unbound} settles this question.

\begin{corollary}
For any $k\geq 0$, $B_k$-CPG is strictly contain in $B_{k+1}$-CPG, even within the class of planar graphs.
\end{corollary}

Note finally that Theorem \ref{thm:unbound} implies that the class of planar CPG graphs has an unbounded bend number (the bend number of a graph class $\mathcal{G}$ is the smallest $k \geq 0$ such that $\mathcal{G} \subseteq B_k$-CPG).


\section{Preliminaries}

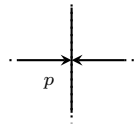
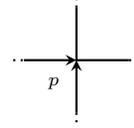
\begin{figure}
\centering
\begin{subfigure}[b]{.45\textwidth}
\centering
\begin{tikzpicture}[scale=.7]
\node at (0,0) (p)  [label=below left:{\scriptsize $p$}]  {};
\draw[thick,-,>=stealth] (0,1)--(0,-1);
\draw[thick,dotted] (0,1.2)--(0,-1.2);
\draw[thick,<-,>=stealth] (0,0)--(1,0);
\draw[thick,dotted] (1,0)--(1.2,0);
\draw[thick,<-,>=stealth] (0,0)--(-1,0);
\draw[thick,dotted] (-1,0)--(-1.2,0);
\end{tikzpicture}
\caption{A grid-point $p$ of type II.a.}
\label{fig:typeIIa}
\end{subfigure}
\begin{subfigure}[b]{.45\textwidth}
\centering
\begin{tikzpicture}[scale=.7]
\node at (3,0) (p)  [label=below left:{\scriptsize $p$}]  {};
\draw[thick,-,>=stealth] (3,1)--(3,0)--(4,0);
\draw[thick,dotted] (3,1)--(3,1.2);
\draw[thick,dotted] (4,0)--(4.2,0);
\draw[thick,<-,>=stealth] (3,0)--(3,-1);
\draw[thick,dotted] (3,-1)--(3,-1.2);
\draw[thick,<-,>=stealth] (3,0)--(2,0);
\draw[thick,dotted] (1.8,0)--(2,0);
\end{tikzpicture}
\caption{A grid-point $p$ of type II.b.}
\label{fig:typeIIb}
\end{subfigure}
\caption{Two types of grid-points (the endpoints are marked by an arrow).}
\end{figure}

Let $G=(V(G),E(G))$ be a CPG graph and $\mathcal{R} = (\mathcal{G},\mathcal{P})$ be a CPG representation of $G$. The path in $\mathcal{P}$ representing some vertex $u \in V(G)$ is denoted by $P_u$. An \textit{interior point} of a path $P$ is a point belonging to $P$ and different from its endpoints; the \textit{interior} of $P$ is the set of all its interior points. A grid-point $p$ is of \textit{type II.a} if it is an endpoint of two paths and an interior point, different from a bendpoint, of a third path~(see Fig. \ref{fig:typeIIa}); a grid-point $p$ is of \textit{type II.b} if it is an endpoint of two paths and a bendpoint of a third path~(see Fig. \ref{fig:typeIIb}).

\section{Proof of Theorem \ref{thm:unbound}}

\begin{figure}[hb]
\centering
\begin{subfigure}{.65\linewidth}
\centering
\begin{tikzpicture}[scale=0.55]
\fill[green!20!white] (-4.75,0) ellipse (1.25cm and 1cm);
\fill[green!20!white] (-2.25,0) ellipse (1.25cm and 1cm);
\fill[green!20!white] (4.75,0) ellipse (1.25cm and 1cm);
\node at (-4.75,0) {$H_1$};
\node at (-2.25,0) {$H_2$};
\node at (4.75,0) {$H_{19}$};

\node[circ] (a) at (0,3) [label=above:{\tiny $a$}] {};
\node[circ] (b) at (0,-3) [label=below:{\tiny $b$}] {};

\node[circ] (a1) at (-6,0) [label=below left:{\tiny $\alpha_1$}] {};
\node[circ] (a2) at (-3.5,0) [label=below right:{\tiny $\alpha_2$}] {};
\node[circ] (a3) at (-1,0) [label=below right:{\tiny $\alpha_3$}] {};
\node[circ] (a19) at (3.5,0) [label=below left:{\tiny $\alpha_{19}$}] {};
\node[circ] (a20) at (6,0) [label=below right:{\tiny $\alpha_{20}$}] {};

\draw (a) .. controls (-6,2) .. (a1);
\draw (a) .. controls (-3.5,2) .. (a2);
\draw (a) .. controls (-1,2) .. (a3);
\draw (a) .. controls (3.5,2) .. (a19);
\draw (a) .. controls (6,2) .. (a20);
\draw (a) .. controls (-11,4) and (-11,-4) .. (b);
\draw (b) .. controls (-6,-2) .. (a1);
\draw (b) .. controls (-3.5,-2) .. (a2);
\draw (b) .. controls (-1,-2) .. (a3);
\draw (b) .. controls (3.5,-2) .. (a19);
\draw (b) .. controls (6,-2) .. (a20);

\node[scirc] at (1,0) {};
\node[scirc] at (1.25,0) {};
\node[scirc] at (1.5,0) {};

\node[invisible] at (2,-4.8) {};
\end{tikzpicture}
\caption{The planar graph $G_k$.}
\label{fig:overall}
\end{subfigure}
\begin{subfigure}{.3\linewidth}
\centering
\begin{tikzpicture}[scale=0.6]
\node[circ] (ai) at (0,0) [label=left:{\tiny $\alpha_i$}] {};
\node[circ] (ai1) at (4,0) [label=right:{\tiny $\alpha_{i+1}$}] {};

\node[circ] (u1) at (2,3) [label=above right:{\tiny $u_1^i$}] {};
\node[circ] (u2) at (2,2) [label=above right:{\tiny $u_2^i$}] {};
\node[circ] (u3) at (2,1) [label=above right:{\tiny $u_3^i$}] {};
\node[circ] (uk1) at (2,-2) [label=below right:{\tiny $u_{k+1}^i$}] {};
\node[circ] (uk2) at (2,-3) [label=below right:{\tiny $u_{k+2}^i$}] {};

\draw (ai) .. controls (0.5,3) .. (u1);
\draw (ai) .. controls (0.5,2) .. (u2);
\draw (ai) .. controls (0.5,1) .. (u3);
\draw (ai) .. controls (0.5,-2) .. (uk1);
\draw (ai) .. controls (0.5,-3) .. (uk2);
\draw (ai1) .. controls (3.5,3) .. (u1);
\draw (ai1) .. controls (3.5,2) .. (u2);
\draw (ai1) .. controls (3.5,1) .. (u3);
\draw (ai1) .. controls (3.5,-2) .. (uk1);
\draw (ai1) .. controls (3.5,-3) .. (uk2);

\draw[-]
(u1) -- (u2) -- (u3)
(uk1) -- (uk2);

\node[scirc] at (2,0) {};
\node[scirc] at (2,-0.5) {};
\node[scirc] at (2,-1) {};

\end{tikzpicture}
\caption{The gadget $H_i$.}
\label{fig:gadget}
\end{subfigure}
\caption{The construction of $G_k$.}
\label{fig:Gk}
\end{figure}
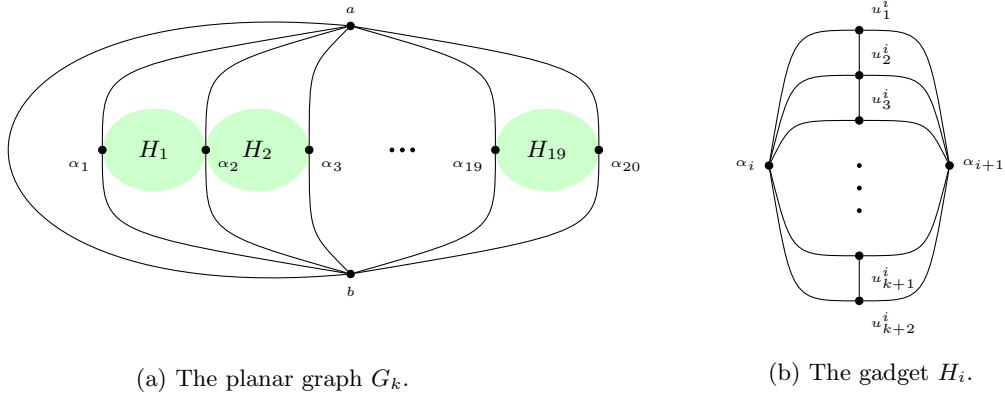

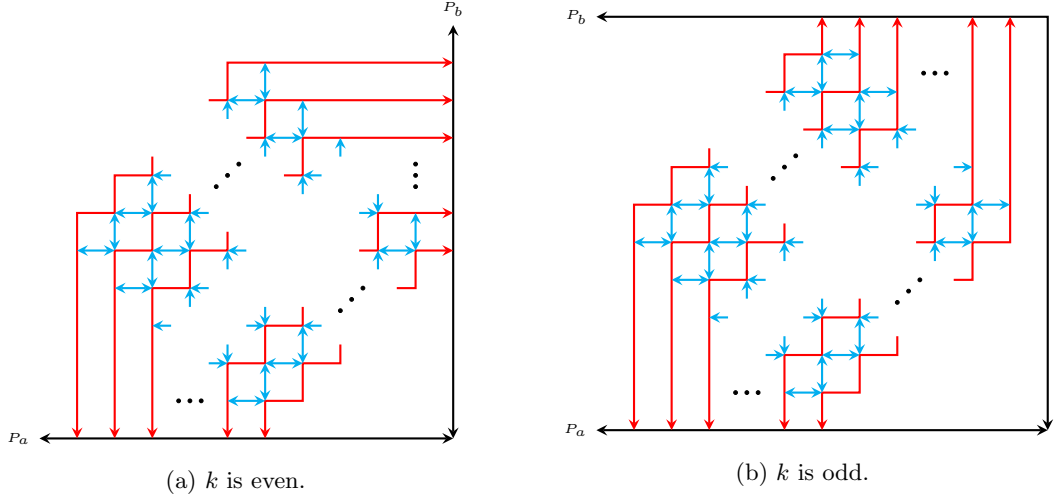
\begin{figure}
\centering
\begin{subfigure}{.45\linewidth}
\centering
\begin{tikzpicture}[scale=0.5]
\draw[thick,<->,>=stealth] (12,-1) -- (1,-1) node[left] {\tiny $P_a$};
\draw[thick,<->,>=stealth] (12,-1) -- (12,10) node[above] {\tiny $P_b$};

\draw[red,thick,<-,>=stealth] (2,-1) -- (2,5) -- (3,5) -- (3,6) -- (4,6) -- (4,6.5);
\draw[red,thick,<-,>=stealth] (3,-1) -- (3,4) -- (4,4) -- (4,5) -- (5,5) -- (5,5.5);
\draw[red,thick,<-,>=stealth] (4,-1) -- (4,3) -- (5,3) -- (5,4) -- (6,4) -- (6,4.5);
\draw[red,thick,<-,>=stealth] (6,-1) -- (6,1) -- (7,1) -- (7,2) -- (8,2) -- (8,2.5);
\draw[red,thick,<-,>=stealth] (7,-1) -- (7,0) -- (8,0) -- (8,1) -- (9,1) -- (9,1.5);

\draw[red,thick,->,>=stealth] (5.5,8) -- (6,8) -- (6,9) -- (12,9);
\draw[red,thick,->,>=stealth] (6.5,7) -- (7,7) -- (7,8) -- (12,8);
\draw[red,thick,->,>=stealth] (7.5,6) -- (8,6) -- (8,7) -- (12,7);
\draw[red,thick,->,>=stealth] (9.5,4) -- (10,4) -- (10,5) -- (12,5);
\draw[red,thick,->,>=stealth] (10.5,3) -- (11,3) -- (11,4) -- (12,4);

\draw[cyan,thick,<->,>=stealth] (2,4) -- (3,4);
\draw[cyan,thick,<->,>=stealth] (3,5) -- (3,4);
\draw[cyan,thick,<->,>=stealth] (3,5) -- (4,5);
\draw[cyan,thick,<->,>=stealth] (4,6) -- (4,5);
\draw[cyan,thick,<-,>=stealth] (4,6) -- (4.5,6);
\draw[cyan,thick,<-,>=stealth] (6,8) -- (6,7.5);
\draw[cyan,thick,<->,>=stealth] (6,8) -- (7,8);
\draw[cyan,thick,<->,>=stealth] (7,9) -- (7,8);

\draw[cyan,thick,<->,>=stealth] (3,3) -- (4,3);
\draw[cyan,thick,<->,>=stealth] (4,4) -- (4,3);
\draw[cyan,thick,<->,>=stealth] (4,4) -- (5,4);
\draw[cyan,thick,<->,>=stealth] (5,5) -- (5,4);
\draw[cyan,thick,<-,>=stealth] (5,5) -- (5.5,5);
\draw[cyan,thick,<-,>=stealth] (7,7) -- (7,6.5);
\draw[cyan,thick,<->,>=stealth] (7,7) -- (8,7);
\draw[cyan,thick,<->,>=stealth] (8,8) -- (8,7);

\draw[cyan,thick,<-,>=stealth] (4,2) -- (4.5,2);
\draw[cyan,thick,<-,>=stealth] (5,3) -- (5,2.5);
\draw[cyan,thick,<-,>=stealth] (5,3) -- (5.5,3);
\draw[cyan,thick,<-,>=stealth] (6,4) -- (6,3.5);
\draw[cyan,thick,<-,>=stealth] (6,4) -- (6.5,4);
\draw[cyan,thick,<-,>=stealth] (8,6) -- (8,5.5);
\draw[cyan,thick,<-,>=stealth] (8,6) -- (8.5,6);
\draw[cyan,thick,<-,>=stealth] (9,7) -- (9,6.5);

\draw[cyan,thick,->,>=stealth] (5.5,1) -- (6,1);
\draw[cyan,thick,->,>=stealth] (6,1.5) -- (6,1);
\draw[cyan,thick,->,>=stealth] (6.5,2) -- (7,2);
\draw[cyan,thick,->,>=stealth] (7,2.5) -- (7,2);
\draw[cyan,thick,->,>=stealth] (9.5,5) -- (10,5);
\draw[cyan,thick,->,>=stealth] (10,5.5) -- (10,5);

\draw[cyan,thick,<->,>=stealth] (6,0) -- (7,0);
\draw[cyan,thick,<->,>=stealth] (7,1) -- (7,0);
\draw[cyan,thick,<->,>=stealth] (7,1) -- (8,1);
\draw[cyan,thick,<->,>=stealth] (8,2) -- (8,1);
\draw[cyan,thick,<-,>=stealth] (8,2) -- (8.5,2);
\draw[cyan,thick,<-,>=stealth] (10,4) -- (10,3.5);
\draw[cyan,thick,<->,>=stealth] (10,4) -- (11,4);
\draw[cyan,thick,<->,>=stealth] (11,5) -- (11,4);

\node[scirc] at (4.7,0) {};
\node[scirc] at (5,0) {};
\node[scirc] at (5.3,0) {};

\node[scirc] at (11,5.7) {};
\node[scirc] at (11,6) {};
\node[scirc] at (11,6.3) {};

\node[scirc] at (5.7,5.7) {};
\node[scirc] at (6,6) {};
\node[scirc] at (6.3,6.3) {};

\node[scirc] at (9,2.4) {};
\node[scirc] at (9.3,2.7) {};
\node[scirc] at (9.6,3) {};
\end{tikzpicture}
\caption{$k$ is even.}
\end{subfigure}
\begin{subfigure}{.45\linewidth}
\centering
\begin{tikzpicture}[scale=0.5]
\draw[thick,<->,>=stealth] (13,-1) -- (1,-1) node[left] {\tiny $P_a$};
\draw[thick,<->,>=stealth] (13,-1) -- (13,10) -- (1,10) node[left] {\tiny $P_b$};

\draw[red,thick,<-,>=stealth] (2,-1) -- (2,5) -- (3,5) -- (3,6) -- (4,6) -- (4,6.5);
\draw[red,thick,<-,>=stealth] (3,-1) -- (3,4) -- (4,4) -- (4,5) -- (5,5) -- (5,5.5);
\draw[red,thick,<-,>=stealth] (4,-1) -- (4,3) -- (5,3) -- (5,4) -- (6,4) -- (6,4.5);
\draw[red,thick,<-,>=stealth] (6,-1) -- (6,1) -- (7,1) -- (7,2) -- (8,2) -- (8,2.5);
\draw[red,thick,<-,>=stealth] (7,-1) -- (7,0) -- (8,0) -- (8,1) -- (9,1) -- (9,1.5);

\draw[red,thick,->,>=stealth] (5.5,8) -- (6,8) -- (6,9) -- (7,9) -- (7,10);
\draw[red,thick,->,>=stealth] (6.5,7) -- (7,7) -- (7,8) -- (8,8) -- (8,10);
\draw[red,thick,->,>=stealth] (7.5,6) -- (8,6) -- (8,7) -- (9,7) -- (9,10);
\draw[red,thick,->,>=stealth] (9.5,4) -- (10,4) -- (10,5) -- (11,5) -- (11,10);
\draw[red,thick,->,>=stealth] (10.5,3) -- (11,3) -- (11,4) -- (12,4) -- (12,10);

\draw[cyan,thick,<->,>=stealth] (2,4) -- (3,4);
\draw[cyan,thick,<->,>=stealth] (3,5) -- (3,4);
\draw[cyan,thick,<->,>=stealth] (3,5) -- (4,5);
\draw[cyan,thick,<->,>=stealth] (4,6) -- (4,5);
\draw[cyan,thick,<-,>=stealth] (4,6) -- (4.5,6);
\draw[cyan,thick,<-,>=stealth] (6,8) -- (6,7.5);
\draw[cyan,thick,<->,>=stealth] (6,8) -- (7,8);
\draw[cyan,thick,<->,>=stealth] (7,9) -- (7,8);
\draw[cyan,thick,<->,>=stealth] (7,9) -- (8,9);

\draw[cyan,thick,<->,>=stealth] (3,3) -- (4,3);
\draw[cyan,thick,<->,>=stealth] (4,4) -- (4,3);
\draw[cyan,thick,<->,>=stealth] (4,4) -- (5,4);
\draw[cyan,thick,<->,>=stealth] (5,5) -- (5,4);
\draw[cyan,thick,<-,>=stealth] (5,5) -- (5.5,5);
\draw[cyan,thick,<-,>=stealth] (7,7) -- (7,6.5);
\draw[cyan,thick,<->,>=stealth] (7,7) -- (8,7);
\draw[cyan,thick,<->,>=stealth] (8,8) -- (8,7);
\draw[cyan,thick,<->,>=stealth] (8,8) -- (9,8);

\draw[cyan,thick,<-,>=stealth] (4,2) -- (4.5,2);
\draw[cyan,thick,<-,>=stealth] (5,3) -- (5,2.5);
\draw[cyan,thick,<-,>=stealth] (5,3) -- (5.5,3);
\draw[cyan,thick,<-,>=stealth] (6,4) -- (6,3.5);
\draw[cyan,thick,<-,>=stealth] (6,4) -- (6.5,4);
\draw[cyan,thick,<-,>=stealth] (8,6) -- (8,5.5);
\draw[cyan,thick,<-,>=stealth] (8,6) -- (8.5,6);
\draw[cyan,thick,<-,>=stealth] (9,7) -- (9,6.5);
\draw[cyan,thick,<-,>=stealth] (9,7) -- (9.5,7);

\draw[cyan,thick,->,>=stealth] (5.5,1) -- (6,1);
\draw[cyan,thick,->,>=stealth] (6,1.5) -- (6,1);
\draw[cyan,thick,->,>=stealth] (6.5,2) -- (7,2);
\draw[cyan,thick,->,>=stealth] (7,2.5) -- (7,2);
\draw[cyan,thick,->,>=stealth] (9.5,5) -- (10,5);
\draw[cyan,thick,->,>=stealth] (10,5.5) -- (10,5);
\draw[cyan,thick,->,>=stealth] (10.5,6) -- (11,6);

\draw[cyan,thick,<->,>=stealth] (6,0) -- (7,0);
\draw[cyan,thick,<->,>=stealth] (7,1) -- (7,0);
\draw[cyan,thick,<->,>=stealth] (7,1) -- (8,1);
\draw[cyan,thick,<->,>=stealth] (8,2) -- (8,1);
\draw[cyan,thick,<-,>=stealth] (8,2) -- (8.5,2);
\draw[cyan,thick,<-,>=stealth] (10,4) -- (10,3.5);
\draw[cyan,thick,<->,>=stealth] (10,4) -- (11,4);
\draw[cyan,thick,<->,>=stealth] (11,5) -- (11,4);
\draw[cyan,thick,<->,>=stealth] (11,5) -- (12,5);

\node[scirc] at (4.7,0) {};
\node[scirc] at (5,0) {};
\node[scirc] at (5.3,0) {};

\node[scirc] at (9.7,8.5) {};
\node[scirc] at (10,8.5) {};
\node[scirc] at (10.3,8.5) {};

\node[scirc] at (5.7,5.7) {};
\node[scirc] at (6,6) {};
\node[scirc] at (6.3,6.3) {};

\node[scirc] at (9,2.4) {};
\node[scirc] at (9.3,2.7) {};
\node[scirc] at (9.6,3) {};
\end{tikzpicture}
\caption{$k$ is odd.}
\end{subfigure}
\caption{A $(k+1)$-bend CPG representation of $G_k$.}
\label{fig:representation}
\end{figure}

We show that the planar graph $G_k$, with $k \geq 0$, depicted in Fig.~\ref{fig:Gk} is in $B_{k+1}$-CPG $\setminus B_k$-CPG. We refer to the vertices $\alpha_i$, for $1 \leq i \leq 20$, as the \textit{secondary vertices}, and to the vertices $u_j^i$, for $1 \leq j \leq k+2$ and a given $1 \leq i \leq 19$, as the \textit{$(i,i+1)$-sewing vertices}. In Fig.~\ref{fig:representation} is given a $(k+1)$-bend CPG representation of $G_k$ (where the blue paths correspond to sewing vertices and the red paths correspond to secondary vertices). We next prove that in any CPG representation of $G_k$, there exists a path with at least $k+1$ bends.

Let $\mathcal{R} = (\mathcal{G},\mathcal{P})$ be a CPG representation of $G_k$. A path in $\mathcal{P}$ corresponding to a secondary vertex (resp. an $(i,i+1)$-sewing vertex) is called a \textit{secondary path} (resp. an \textit{$(i,i+1)$-sewing path}). A secondary path $P_{\alpha_i}$ is said to be \textit{pure} if no endpoint of $P_a$ or $P_b$ belongs to $P_{\alpha_i}$. We then have the following easy observation.

\begin{observation}
\label{pureendpoints}
If a path $P_{\alpha_i}$ is pure, then one endpoint of $P_{\alpha_i}$ belongs to $P_a$ and the other endpoint belongs to $P_b$.
\end{observation}

\begin{Claim}
\label{bendpoint}
Let $P_{\alpha_i}$ and $P_{\alpha_{i+1}}$ be two pure paths and let $u$ and $v$ be two $(i,i+1)$-sewing vertices such that $uv \in E(G_k)$. If a grid-point $x$ belongs to $P_u \cap P_v$, then $x$ corresponds to an endpoint of both $P_u$ and $P_v$, and a bendpoint of either $P_{\alpha_i}$ or $P_{\alpha_{i+1}}$.
\end{Claim}

It follows from Observation \ref{pureendpoints} and the fact that $u$ is non-adjacent to both $a$ and $b$, that no endpoint of $P_{\alpha_i}$ or $P_{\alpha_{i+1}}$ belongs to $P_u$. Consequently, one endpoint of $P_u$ belongs to $P_{\alpha_i}$ and the other endpoint belongs to $P_{\alpha_{i+1}}$. We conclude similarly for $P_v$. By definition, $x$ corresponds to an endpoint of at least one of $P_u$ and $P_v$, which implies that $x$ belongs to $\mathring{P_{\alpha_i}}$ or $\mathring{P_{\alpha_{i+1}}}$. But then, $x$ must be an endpoint of both $P_u$ and $P_v$; in particular, $x$ is a grid-point of type either II.a or II.b. Without loss of generality, we may assume that $x \in \mathring{P_{\alpha_i}}$. We denote by $y_u$ (resp. $y_v$) the endpoint of $P_u$ (resp. $P_v$) belonging to $P_{\alpha_{i+1}}$. Now, suppose to the contrary that $x$ is of type II.a. The union of $P_u$, $P_v$ and the portion of $P_{\alpha_{i+1}}$ between $y_u$ and $y_v$ defines a closed curve $\mathcal{C}$, which divides the plane into two regions. Since $P_a$ and $P_b$ touch neither $P_u$, $P_v$ nor $\mathring{P_{\alpha_{i+1}}}$ (recall that $P_{\alpha_{i+1}}$ is pure), $P_a$ and $P_b$ lie entirely in one of those regions; and, as $a$ and $b$ are adjacent, $P_a$ and $P_b$ in fact belong to the same region. On the other hand, since one endpoint of $P_u$ (resp. $P_v$) belongs to $P_{\alpha_i}$ while the other endpoint belongs to $P_{\alpha_{i+1}}$, and both endpoints of $P_{\alpha_i}$ are in $P_a \cup P_b$, it follows that $x \in \mathcal{C}$ is the only contact point between $P_u$ (resp. $P_v$) and $P_{\alpha_i}$; but $\alpha_i$ and $\alpha_{i+1}$ being non-adjacent, this implies that $P_{\alpha_i}$ crosses $\mathcal{C}$ only once and has therefore one endpoint in each region. But both endpoints of $P_{\alpha_i}$ belong to $P_a \cup P_b$ which contradicts the fact that $P_a$ and $P_b$ lie in the same region. Hence, $x$ is of type II.b which concludes the proof.~$\diamond$

\bigskip

\begin{Claim}
\label{bends}
If two paths $P_{\alpha_i}$ and $P_{\alpha_{i+1}}$ are pure, then one of them contains at least $\lfloor \frac{k+1}{2} \rfloor$ bends and the other one contains at least $\lceil \frac{k+1}{2} \rceil$ bends. Moreover, all of those bendpoints belong to \textit{(i,i+1)}-sewing paths.
\end{Claim}

For all $1 \leq j \leq k+1$, consider a point $x_j \in P_{u_j^i} \cap P_{u_{j+1}^i}$. It follows from Claim \ref{bendpoint} that $x_j$ is a bendpoint of either $P_{\alpha_i}$ or $P_{\alpha_{i+1}}$. Since $x_j$ and $x_{j+1}$ are the endpoints of $P_{u_{j+1}^i}$, one belongs to $P_{\alpha_i}$ while the other belongs to $P_{\alpha_{i+1}}$. Therefore, $\{x_j, 1 \leq j \leq k+1 \text{ and } (j \text{ mod } 2) = 0\}$ is a subset of one of the considered secondary paths and $\{x_j, 1 \leq j \leq k+1 \text{ and } (j \text{ mod } 2) = 1\}$ is a subset of the other secondary path. $\diamond$

\bigskip

Finally, we claim that there exists an index $1 \leq j \leq 17$ such that $P_{\alpha_j}$, $P_{\alpha_{j+1}}$, $P_{\alpha_{j+2}}$ and $P_{\alpha_{j+3}}$ are all pure. Indeed, if it weren't the case, then there would be at least $\lfloor 20/4 \rfloor = 5$ secondary paths that are not pure; but at most $4$ secondary paths can contain endpoints of $P_a$ or $P_b$, a contradiction. It now follows from Claim \ref{bends} that $P_{\alpha_{j+1}}$ has at least $\lfloor \frac{k+1}{2} \rfloor$ bends (which belong to $(j,j+1)$-sewing paths), and that $P_{\alpha_{j+2}}$ has at least $\lfloor \frac{k+1}{2} \rfloor$ bends (which belong to $(j+2,j+3)$-sewing paths). Furthermore, either $P_{\alpha_{j+1}}$ or $P_{\alpha_{j+2}}$ has at least $\lceil \frac{k+1}{2} \rceil$ bends which are endpoints of $(j+1,j+2)$-sewing paths. Either way, there is a path with at least $\lfloor \frac{k+1}{2} \rfloor + \lceil \frac{k+1}{2} \rceil = k+1$ bends, which concludes the proof of Theorem \ref{thm:unbound}.


\section{Conclusion}

In this note, we prove that the class of planar CPG graphs is not included in any $B_k$-CPG, for $k \geq 0$. More specifically, we show that for any $k \geq 0$, there exists a planar graph which is $B_{k+1}$-CPG but not $B_k$-CPG. As a consequence, we also obtain that $B_k$-CPG $\subsetneq$ $B_{k+1}$-CPG for any $k \geq 0$.

\bibliography{biblio}

\begin{thebibliography}{1}

\bibitem{cpg}
Z.~Deniz, E.~Galby, A.~Munaro, and B.~Ries.
\newblock On contact graphs of paths on a grid.
\newblock {\em CoRR}, abs/1803.03468, 2018.

\end{thebibliography}
\end{document}